\documentclass{jpsj3}  %
\usepackage{txfonts}

  \title{Ginzburg-Landau Theory for Flux Phase and Superconductivity 
in $t-J$ Model
}
 
\author{Kazuhiro Kuboki \thanks{kuboki@kobe-u.ac.jp}}
\inst{Department of Physics, Kobe University, Kobe 657-8501, Japan
} 

\abst{
Ginzburg-Landau (GL) equations and GL free energy for 
flux phase and superconductivity are derived microscopically 
from the $t-J$ model on a square lattice.
Order parameter (OP) for the flux phase has  direct coupling 
to a magnetic field, in contrast to the superconducting OP 
which has minimal coupling to a vector potential. 
Therefore, when the flux phase OP has unidirectional spatial 
variation, staggered currents would flow in a perpendicular 
direction. The derived GL theory can be used for 
various problems in high-$T_c$ cuprate superconductors, 
e.g., states near a surface or impurities, 
and the effect of an external magnetic field. 
Since the GL theory derived microscopically directly reflects 
the electronic structure of the system, e.g., the shape 
of the Fermi surface that changes with doping,  
it can provide more useful information than that from  
phenomenological GL theories.
}

\begin{document}
\maketitle

\newpage

\section{Introduction}

Spontaneous violation of time-reversal symmetry ${\cal T}$ 
in superconductors has been discussed intensively. 
Especially, in high-$T_c$ cuprate superconductors, 
a sign of spontaneous ${\cal T}$ violation was detected two decades ago.
Covington {\it et al.} observed the peak splitting of zero bias 
conductance in ab-oriented YBCO/insulator/Cu 
junction,\cite{Coving}  and this has been interpreted as 
evidence of ${\cal T}$ violation 
induced by  a second superconducting (SC) 
order parameter (OP) near the surface, with the symmetry 
different from that in the bulk.\cite{Matsu1,Matsu2,Fogel}   
In the literature, surface (or interface) ${\cal T}$-breaking states 
with $(d \pm is)$-, $(d\pm id')$-, and $(d \pm ig)$-wave 
symmetry have been 
discussed.\cite{SBR,KK,Matsu1,Matsu2,Fogel,SigRev,Kirt}  
However, existence of spontaneous current which would flow 
along the surface  and an accompanying magnetic field 
is still controversial.\cite{Carmi,msr}

The present author has studied (110) surface states of high-$T_C$ 
cuprates based on the Bogoliubov-de Gennes (BdG) method 
by employing single-layer and bilayer $t-J$ models, and found 
that a different kind of ${\cal T}$-breaking surface state, flux phase,  
can occur.\cite{KK1,KK2,KK3,KK4} 
Flux phase is a mean-field solution to the $t-J$ model in which 
staggered currents flow and the flux penetrates a plaquette in 
a square lattice.\cite{Affleck} 
(The $d$-density wave state, which has been introduced in 
a different context, have similar properties.\cite{cha}) 
Mean-field\cite{Zhang,Ubb,Hamada,Zhao,Bejas} and 
variational Monte Carlo\cite{Yoko1,Liang,TKLee,Ivanov} 
calculations have shown that 
free energy of the flux state is higher than that of 
a $d_{x^2-y^2}$-wave SC state except very near half filling, 
so that it is only a next-to-leading (metastable) 
state in uniform systems.
(Inclusion of nearest-neighbor repulsive interactions to the model 
may lead to the appearance of the flux phase for small 
doping rates.\cite{Zhou})
Besides the $t-J$ model, variational Monte Carlo calculations 
for the Hubbard model have shown that coexistence of flux phase 
and superconductivity cannot occur in uniform systems,\cite{Yoko2} 
and this feature is consistent with the results for the $t-J$ model.
However, near the (110) surface, the $d_{x^2-y^2}$-wave SC state 
is strongly suppressed, and the flux phase may arise 
locally leading to a ${\cal T}$-breaking surface state. 
Since the spontaneous current in this state is a staggered one, 
accompanying magnetic fields will be smaller compared to those 
induced in surface states with the second SCOP.

The peak splitting of zero bias conductance was observed in systems  
composed of YBCO cuprates that has two CuO$_2$ planes in a unit cell. 
In the case of a bilayer $t-J$ model, which is a model to describe 
bilayer cuprates such as YBCO, 
there may be two types of flux phase in which the directions of 
the flux in two layers are the same or opposite.
Mean-field approximation (MFA) and BdG 
calculations have shown that the latter state has lower energy 
than that of the former.\cite{KK3,KK4}
Then the spontaneous currents and magnetic fields  
in two layers have opposite signs, and so the observed field 
near the surface will be smaller compared to the single-layer
systems. 
The theoretically obtained magnetic field is, however, 
still comparable or larger than the upper bound set by experiments. 
In order to determine whether or not the peak splitting can be 
explained in terms of the flux state, 
we have to evaluate the magnetic field strength in a 
self-consistent manner, 
which is difficult to carry out in BdG calculations. 

In this paper, we derive Ginzburg-Landau (GL) equations 
and GL free energy microscopically from the two-dimensional 
$t-J$ model on a square lattice, in order to provide 
a tool to investigate the distribution of magnetic fields 
more accurately. 
The method of deriving GL equations is essentially the same 
as that used in previous studies which treat the coexistence of 
magnetic order and superconductivity in the extended 
Hubbard model\cite{KKGL1} and the $t-J$ model.\cite{KKGL2}
The resulting GL theory can be used for various problems, 
e.g., investigation of the magnetic field distribution 
near the surface, states near impurities, 
and the effect of an external magnetic field. 
Although the GL theory is reliable only qualitatively
except near $T_C$, it can give simple and intuitive description of
the coexistence and competition of multiple OPs. Thus, it is
complementary to more sophisticated methods such as the
BdG and quasiclassical Green's function theory. 

This paper is organized as follows. In Sect. 2 we present the model 
and treat it by a mean-field approximation. GL equations and GL free 
energy are derived in Sect. 3. Sect. 4 is devoted to summary.

\section{Model and Mean-Field Approximation}

We consider the $t-J$ model on a square lattice 
whose Hamiltonian is given by 
\begin{eqnarray}
\displaystyle 
H = \displaystyle -\sum_{j,\ell,\sigma} 
t_{j\ell} {\tilde c}^\dagger_{j\sigma} {\tilde c}_{\ell\sigma}
 +J\sum_{\langle j,\ell\rangle} {\bf S}_j \cdot {\bf S}_\ell, 
\end{eqnarray}
where the transfer integrals $t_{j\ell}$ are finite for the first-  ($t$), 
second-  ($t'$), and third-nearest-neighbor bonds ($t''$), or zero otherwise.  
$J$ is the antiferromagnetic superexchange
 interaction, and $\langle j,\ell \rangle$ denotes nearest-neighbor bonds.\cite{Ogata} 
 The magnetic field is taken into account using the Peierls phase 
$\phi_{j,\ell} \equiv \frac{\pi}{\phi_0} \int_j^\ell {\bf A}\cdot d{\bf l}$, 
with ${\bf A}$ and $\phi_0 = \frac{h}{2e}$ being the vector potential 
and  flux quantum, respectively. 
${\tilde c}_{j\sigma}$ is the electron operator in Fock space without double occupancy, and we treat this condition 
using the slave-boson method\cite{Ogata,Zou,Lee}   
by writing ${\tilde c}_{j\sigma}=b_j^\dagger f_{j\sigma}$ 
under the local constraint 
$\sum_{\sigma}f_{j\,\sigma}^\dagger f_{j\,\sigma} 
+ b_j^\dagger b_j = 1$ 
at every $j$ site. Here $f_{j\sigma}$ ($b_j$) is a fermion (boson) operator  
that carries spin $\sigma$ (charge $e$); the fermions (bosons) are frequently 
referred to as spinons (holons). 
The spin operator is expressed as 
$ {\bf S}_j  = \frac{1}{2}\sum_{\alpha,\beta}
f^\dagger_{j\alpha} {\bf \sigma}_{\alpha\beta}f_{j\beta}$. 

We decouple the Hamiltonian in the following 
manner.\cite{Kotliar,Suzumura} 
The bond OPs 
$ \sum_\sigma \langle f^\dagger_{j\sigma}f_{\ell\sigma} \rangle$ 
and $ \langle b^\dagger_j b_\ell\rangle$ 
are introduced, and we denote 
$\chi_{j,\ell} \equiv  
\sum_\sigma \langle f^\dagger_{j\sigma}f_{\ell\sigma} \rangle$
for nearest-neighbor bonds. 
Although the bosons are not condensed in purely two-dimensional systems  
at finite temperature ($T$), 
they are almost condensed at a low $T$ and for 
finite carrier doping ($\delta \gtrsim 0.05$, $\delta$ being the 
doping rate). Since we are interested in the low temperature region
 ($T \lesssim 10^{-2}J \sim 10$K) and the doping rate 
$\delta \gtrsim 0.05$, we treat holons as Bose condensed.
Hence, we approximate $ \langle b_j \rangle \sim \sqrt{\delta}$ and 
$ \langle b^\dagger_j b_\ell \rangle \sim \delta$, 
and replace the local constraint with a global one, 
$\frac{1}{N}\sum_{j,\sigma} \langle f^\dagger_{j\sigma} f_{j\sigma}\rangle 
= 1-\delta$, where $N$ is the total number of lattice sites.
This procedure amounts to renormalizing the transfer integrals 
by multiplying $\delta$, 
i.e., $t \to t\delta$, {\it etc.}, 
and rewriting ${\tilde c}_{j\sigma}$ as $f_{j\sigma}$. 
In a qualitative sense, this approach is equivalent to 
the renormalized mean-field theory of Zhang {\it et al.}\cite{Zhang2} 
(Gutzwiller approximation). 
The spin-singlet resonating-valence-bond (RVB) OP on the bond 
$\langle j,\ell\rangle$ is given by  
$\Delta_{j,\ell} = \langle f_{j\uparrow} f_{\ell\downarrow}
-f_{j\downarrow} f_{\ell\uparrow} \rangle/2$. 
Under the assumption of the Bose condensation of holons, 
$\Delta_{j,\ell}$ is equivalent to the SCOP. 

With the above definitions of the OPs, the mean-field Hamiltonian  
is written as 
\begin{eqnarray}
H_{MFA} =  & \displaystyle -\sum_{j,\sigma}\Big[
\sum_{\delta_1=\pm {\hat x},\pm {\hat y}} 
\Big(t\delta e^{i\phi_{j+\delta_1,j}} 
+ \frac{3J}{8}\chi_{j,j+\delta_1}\Big)  
f^\dagger_{j+\delta_1,\sigma}f_{j\sigma} \nonumber  \\
+& \displaystyle  t'\delta\sum_{\delta_2=\pm {\hat x} \pm {\hat y}} 
e^{i\phi_{j+\delta_2,j}} f^\dagger_{j+\delta_2,\sigma}f_{j\sigma}    
+ t''\delta\sum_{\delta_3=\pm 2{\hat x}, \pm 2{\hat y}}
 e^{i\phi_{j+\delta_3,j}}  f^\dagger_{j+\delta_3,\sigma}f_{j\sigma}  
 +  \mu f^\dagger_{j\sigma}f_{j\sigma} \Big] \nonumber  \\
 + & \displaystyle  \frac{3J}{8}\sum_j 
\sum_{\delta_1= \pm {\hat x},\pm {\hat y}}
\Big[\Delta_{j,j+\delta_1}\big(f^\dagger_{j\uparrow}f^\dagger_{j
+\delta_1\downarrow} 
-f^\dagger_{j\downarrow}f^\dagger_{j+\delta_1\uparrow}\big) + h.c.\Big] 
+ E_0,  
\end{eqnarray}
where
\begin{eqnarray}
E_0 =  & \displaystyle \frac{3J}{2}\sum_j
\sum_{\delta_1={\hat x},{\hat y}}\Big(|\Delta_{j,j+\delta_1}|^2 
+ \frac{1}{4} | \chi_{j,j+\delta_1}|^2\Big),  
\end{eqnarray}
and $\mu$ is the chemical potential.
We divide $\chi_{j,\ell}$ into two parts
\begin{eqnarray}
\displaystyle \chi_{j,\ell} = \chi_0 + Z_{j,\ell}, 
\end{eqnarray}
where $\chi_0$ is real and uniform  in space, 
while $Z_{j,\ell}$ may be complex and describe the flux phase 
as we will see in the following. 

Since the onset temperature of $\chi_0$ is much higher than 
that for superconductivity ($T_C$) and the bare transition 
temperature of the flux phase,  
we treat only $\Delta$ and $Z$ as the GL-expansion 
parameters, and determine $\chi_0$ using the usual MFA.
The self-consistency equations for $\chi_0$ and $\mu$ 
in the absence of $\Delta$, $Z$, and ${\bf A}$ are given as 
\begin{eqnarray}
\displaystyle \chi_0 =  \frac{1}{N}\sum_p (\cos p_x+\cos p_y)f(\xi_p) , \ \ 
\delta=  1 - \frac{2}{N}\sum_pf(\xi_p), 
\end{eqnarray}
where  $f$ is the Fermi distribution function, and 
\begin{eqnarray}
\displaystyle \xi_p = -\Big(2t\delta + \frac{3J}{4}\chi_0\Big)(\cos p_x+\cos p_y) 
 -4t'\delta\cos p_x \cos p_y - 2t''\delta(\cos 2p_x+\cos 2p_y) -\mu. 
\end{eqnarray}
We set  the lattice constant to be unity.

\section{Ginzburg-Landau Equations and Free Energy}
In this section we derive the GL equations and GL free energy. 
The Gor'kov equations for normal and anomalous Green's functions, 
respectively defined as, 
$G(j,\ell,\tau) \equiv   
-\langle T_\tau f_{j\uparrow}(\tau)f_{\ell\uparrow}^\dagger\rangle$
and $F^\dagger(j,\ell,\tau) \equiv  
-\langle T_\tau f_{j\downarrow}^\dagger(\tau)
f_{\ell\uparrow}^\dagger\rangle$, 
can be derived by a standard procedure,\cite{Gorkov,KKGL1,KKGL2} 
\begin{eqnarray}
& \displaystyle  (i\epsilon_n + \mu) G(j,\ell,i\epsilon_n) 
+ \sum_{\delta_1=\pm{\hat x},\pm{\hat y}} 
\Big(t\delta e^{i\phi_{j,j+\delta_1}}
+ \frac{3J}{8}\chi_{j+\delta_1.j}\Big) G(j+\delta_1,\ell,i\epsilon_n) \nonumber \\
& \displaystyle + \sum_{\delta_2=\pm{\hat x}\pm{\hat y}} 
t'\delta e^{i\phi_{j,j+\delta_2}}G(j+\delta_2,\ell,i\epsilon_n) 
+ \sum_{\delta_3=\pm 2{\hat x}, \pm 2{\hat y}} t''\delta e^{i\phi_{j,j+\delta_3}}
G(j+\delta_3,\ell,i\epsilon_n) \nonumber  \\
& \displaystyle - \frac{3J}{4} \sum_{\delta_1=\pm{\hat x},\pm{\hat y}}
\Delta_{j,j+\delta_1} F^\dagger(j+\delta_1.\ell,i\epsilon_n) = \delta_{jl}, \\
& \displaystyle (i\epsilon_n - \mu) F^\dagger(j,\ell,i\epsilon_n) 
- \sum_{\delta_1=\pm{\hat x},\pm{\hat y}} 
\Big(t\delta e^{i\phi_{j+\delta_1,j}}
+ \frac{3J}{8}\chi_{j,j+\delta_1}\Big) F^\dagger(j+\delta_1,\ell,i\epsilon_n) 
\nonumber \\
& \displaystyle - \sum_{\delta_2=\pm{\hat x}\pm{\hat y}} 
t'\delta e^{i\phi_{j+\delta_2,j}} F^\dagger(j+\delta_2,\ell,i\epsilon_n) 
- \sum_{\delta_3=\pm 2{\hat x}, \pm 2{\hat y}} t''\delta e^{i\phi_{j+\delta_3,j}}
F^\dagger(j+\delta_3,\ell,i\epsilon_n) \nonumber \\
& \displaystyle - \frac{3J}{4} \sum_{\delta_1=\pm{\hat x},\pm{\hat y}}
\Delta^*_{j,j+\delta_1} G(j+\delta_1,\ell,i\epsilon_n) = 0, 
\end{eqnarray}
where $\epsilon_n$ is a fermionic Matsubara frequency. 
These equations can be combined as 
\begin{eqnarray}
\displaystyle G(j,\ell,i\epsilon_n) = 
& \displaystyle {\tilde G}_0(j,\ell,i\epsilon_n) 
+ \frac{3J}{4} \sum_k\sum_{\delta_1=\pm{\hat x},\pm{\hat y}}
{\tilde G}_0(j,k,i\epsilon_n) 
\Big[\Delta_{k,k+\delta_1} F^\dagger(k+\delta_1,\ell,i\epsilon_n)  \nonumber  \\
& \displaystyle
-\frac{1}{2} Z_{k+\delta_1,k}G(k+\delta_1,\ell,i\epsilon_n)\Big],  \\
\displaystyle F^\dagger(j,\ell,i\epsilon_n) = 
& \displaystyle - \frac{3J}{4} \sum_k\sum_{\delta_1=\pm{\hat x},\pm{\hat y}}
{\tilde G}_0(k,j,-i\epsilon_n) 
\Big[\Delta^*_{k,k+\delta_1}G(k+\delta_1,\ell,i\epsilon_n) \nonumber  \\ 
& \displaystyle 
+ \frac{1}{2} Z_{k,k+\delta_1} F^\dagger(k+\delta_1,\ell,i\epsilon_n)\Big], 
\end{eqnarray}
where the summation on $k$ is taken over all sites, 
and ${\tilde G}_0(j,\ell,i\varepsilon_n)$  is  the Green's function for
the system without $\Delta$ and $Z$  but with ${\bf A}$. 
${\tilde G}_0(j,\ell,i\varepsilon_n)$ is related to Green's function for the 
system without ${\bf A}$, $G_0$, as 
$ {\tilde G}_0(j,\ell,i\varepsilon_n) 
\sim   G_0(j,\ell,i\varepsilon_n) e^{i\phi_{j,\ell}}$, with  
$G_0(j,\ell,i\varepsilon_n)$ being the Fourier transform of 
$G_0({\bf p},i\varepsilon_n) = 1/(i\varepsilon_n-\xi_p)$.  
In the expression of $\xi_p$,  
$\chi_0$ and $\mu$ determined by Eqs.(5) and (6) 
will be substituted. 

$Z_{j,\ell}$ may have real ($X_{j,\ell}$) and 
imaginary ($Y_{j,\ell}$) parts;  
\begin{eqnarray}
\displaystyle Z_{j,\ell} = X_{j,\ell} + iY_{j,\ell}.
\end{eqnarray}
$X_{j,\ell} $ and $Y_{j,\ell} $ describe the bond-order phase  and the flux phase, 
respectively, and we treat only the latter in this paper. 
The spin-singlet SCOP ($\Delta_{j\ell}$) and $Y_{j\ell}$ 
are expressed in terms of $F^\dagger$ and $G$, respectively, 
\begin{eqnarray}
\displaystyle (\Delta_{j,\ell})^* 
= & \displaystyle - \frac{T}{2} \sum_{\varepsilon_n}
\Big[F^\dagger(j,\ell,i\varepsilon_n) 
+ F^\dagger(\ell,j,i\varepsilon_n)\Big], \\
Y_{j,\ell}  =  & \displaystyle \frac{1}{2i}(\chi_{j,\ell}-\chi_{\ell,j})
= iT \sum_{\varepsilon_n} 
\Big[G(j,\ell,i\varepsilon_n) -G(\ell,j,i\varepsilon_n)\Big].  
\end{eqnarray}
We substitute Eqs.(9) and (10) into Eqs.(12) and (13) iteratively 
and keep terms up to the third order in the OPs. 
The $s$-  ($d$-) wave SCOP $\Delta_s$ ($\Delta_d$) and the OP 
for the flux phase $\Pi$ can be constructed by making linear 
combinations of Eqs.(12) and (13), 
\begin{eqnarray}
 \Delta_s(j)  =& \displaystyle\frac{1}{4} 
 \sum_{\eta=\pm{\hat x},\pm{\hat y}} \Delta_{j,j+\eta} ,  \\ 
 \Delta_d(j)  =& \displaystyle\frac{1}{4} 
 \Big(\sum_{\eta=\pm{\hat x}}  \Delta_{j,j+\eta} 
 - \sum_{\eta=\pm{\hat y}}  \Delta_{j,j+\eta}\Big),  \\
 \Pi(j) =  & \displaystyle  
 \frac{1}{4} 
 \Big(Y_{j+{\hat x},j} + Y_{j+{\hat x}+{\hat y},j+{\hat x}}
+ Y_{j+{\hat y},j+{\hat x}+{\hat y}}+Y_{j,j+{\hat y}}\Big)
 e^{i{\bf Q}\cdot{\bf r_j}},  
\end{eqnarray}
with ${\bf Q} \equiv (\pi,\pi)$. 
Here we define $\Delta_s$ and $\Delta_d$ at a lattice site
${\bf r}_j$, while $\Pi$ is defined at the center of 
the plaquette, 
${\bf {\tilde r}}_j = {\bf r}_j+{\hat x}/2+{\hat y}/2$.  
The latter definition is necessary to get a gauge-invariant 
coupling between $\Pi$ and the vector potential. 
(See Appendix A.)
Assuming that the SCOPs and $\Pi$ are slowly varying, we take a continuum 
limit. Terms linear in the OPs are expanded in powers of derivatives 
up to the second order, and the Peierls factor is also expanded 
in powers of ${\bf A}$. 
Then we get the following GL equations, 
\begin{eqnarray}
& \displaystyle 
\alpha_s \Delta_s + 2\beta_s |\Delta_s|^2\Delta_s 
- K_s (D_x^2+D_y^2) \Delta_s - K_{ds}(D_x^2-D_y^2)\Delta_d \nonumber \\
& \displaystyle + \gamma_1|\Delta_d|^2\Delta_s 
+ 2\gamma_2\Delta_d^2\Delta_s^* + \gamma_{s\Pi} \Delta_s \Pi^2=0, 
\end{eqnarray}
\begin{eqnarray}
& \displaystyle 
\alpha_d \Delta_d + 2\beta_d |\Delta_d|^2\Delta_d 
- K_d (D_x^2+D_y^2) \Delta_d - K_{ds}(D_x^2-D_y^2)\Delta_s \nonumber \\
& \displaystyle + \gamma_1|\Delta_s|^2\Delta_d 
+ 2\gamma_2\Delta_s^2\Delta_d^* + \gamma_{d\Pi} \Delta_d \Pi^2 = 0, 
\end{eqnarray}
\begin{eqnarray}
& \displaystyle \alpha_\Pi \Pi + 2\beta_\Pi \Pi^3 
- K_\Pi (\partial_x^2+\partial_y^2) \Pi
+\gamma_{s\Pi}|\Delta_s|^2\Pi + \gamma_{d\Pi}|\Delta_d|^2\Pi 
\nonumber  \\ 
& \displaystyle 
+\frac{\alpha_0}{2} \cos ({\bf Q}\cdot{\bf r})
(\partial_x A_y - \partial_y A_x) = 0,
\end{eqnarray}
where the coefficients appearing in Eqs.(17)-(19) are given 
in the Appendix B, 
and ${\bf D}$ is the gauge-invariant gradient defined as 
${\bf D} = \nabla + \frac{2\pi i}{\phi_0}{\bf A}$. 

The GL free energy $F$ up to the fourth order in the OPs can be 
obtained from the above GL equations in such a way that the 
variation in $F$ with respect to the OPs reproduce Eqs.(17)-(19). 
The result is, 
\begin{eqnarray}
F = & \displaystyle F_\Delta+ F_\Pi + F_{\Delta\Pi} + F_B, \\
F_\Delta = &\displaystyle  \int d^2{\bf r} \Big[
 \alpha_d |\Delta_d|^2+ \beta_d |\Delta_d|^4 +  K_d |{\bf D} \Delta_d|^2  
 + \alpha_s|\Delta_s|^2 + \beta_s |\Delta_s|^4  + K_s |{\bf D} \Delta_s|^2 
 \nonumber \\
 & \displaystyle + \gamma_1 |\Delta_d|^2|\Delta_s|^2 
 + \gamma_2 \big(\Delta_d^2(\Delta_s^*)^2 + c.c.\big) \nonumber  \\
& + K_{ds} \big((D_x\Delta_d)(D_x\Delta_s)^{*} 
 - (D_y\Delta_d)(D_y\Delta_s)^{*} + c.c. \big)\Big],   \\
F_\Pi = & \displaystyle  \int d^2{\bf r} \Big[
\alpha_\Pi \Pi^2 + \beta_\Pi \Pi^4 + K_\pi(\nabla \Pi)^2 
+ \alpha_0 \cos ({\bf Q}\cdot{\bf r}) 
(\partial_x A_y - \partial_y A_x) \Pi\Big], 
\\
F_{\Delta\Pi} = & \displaystyle  \int d^2{\bf r}
\big(\gamma_{d\Pi } |\Delta_d|^2 
+ \gamma_{s\Pi}|\Delta_s|^2\big)\Pi^2, 
\\
F_B = & \displaystyle \frac{1}{2\mu_0} \int d^3{\bf r} 
 (\nabla \times {\bf A})^2.
\end{eqnarray}
Here, $F_\Delta$ and $F_\Pi$ are the free energy for 
superconductivity and flux phase, respectively, and 
$F_{\Delta\Pi}$ represents the coupling between SCOPs and $\Pi$. 
$F_B$ is the energy for the magnetic field.

The currents are obtained by varying $F$ with respect to 
${\bf A}$, namely, 
$J_\nu = 
- \frac{\partial}{\partial A_\nu}(F_\Delta+F_{\Pi})$: 
\begin{eqnarray}
\displaystyle J_x  =  & \displaystyle 
-\alpha_0  \cos ({\bf Q}\cdot{\bf r})\partial_y \Pi
-\frac{2\pi i}{\phi_0}\Big[K_d(\Delta_d \partial_x \Delta_d^* 
- \Delta_d^s \partial_x \Delta_d) 
+ K_s(\Delta_s \partial_x \Delta_s^*
- \Delta_s^* \partial_x \Delta_s)  \nonumber \\
& \displaystyle + K_{ds}(\Delta_d \partial_x \Delta_s^* 
- \Delta_d^* \partial_x \Delta_s
+ \Delta_s \partial_x \Delta_d^* 
- \Delta_s^* \partial_x \Delta_d)\Big] \nonumber \\
& \displaystyle - 2\Bigg(\frac{2\pi}{\phi_0}\Bigg)^2 A_x 
\Big[K_d|\Delta_d|^2 + K_s|\Delta_s|^2
+ K_{ds}(\Delta_d\Delta_s^* + \Delta_d^+\Delta_s)\Big], \\
\displaystyle J_y  =  & \displaystyle 
\alpha_0  \cos ({\bf Q}\cdot{\bf r})\partial_x \Pi
-\frac{2\pi i}{\phi_0}\Big[K_d(\Delta_d \partial_y \Delta_d^* 
- \Delta_d^s \partial_y \Delta_d) 
+ K_s(\Delta_s \partial_y \Delta_s^*
- \Delta_s^* \partial_y \Delta_s)  \nonumber \\
& \displaystyle - K_{ds}(\Delta_d \partial_y \Delta_s^* 
- \Delta_d^* \partial_y \Delta_s
+ \Delta_s \partial_y \Delta_d^* - \Delta_s^* \partial_y \Delta_d)\Big] 
\nonumber \\
& \displaystyle - 2\Bigg(\frac{2\pi}{\phi_0}\Bigg)^2 A_y 
\Big[K_d|\Delta_d|^2 + K_s|\Delta_s|^2 
- K_{ds}(\Delta_d\Delta_s^* + \Delta_d^+\Delta_s)\Big].
\end{eqnarray}
Equations (17)-(19), together with the Maxwell equation 
$\nabla \times {\bf B} = \mu_0 {\bf J}$,  
determine $\Delta_d$, $\Delta_s$, $\Pi$, 
and ${\bf A}$ self-consistently. 

Equations (25) and (26) show that when the flux phase OP, $\Pi$, has 
unidirectional spatial variation, 
a current would flow in a perpendicular direction. 
For example, near a (110) surface of a $d$-wave superconductor, 
SCOP is suppressed and then $\Pi$ may become finite. 
Induced $\Pi$ should be uniform along the surface and decays 
toward the bulk. 
In this case, staggered currents would flow along the surface, 
and their amplitudes vanish in the bulk. 
Near the surface, spontaneous magnetic fields would appear.   
Their spatial distribution can be determined by solving 
the GL equations numerically, and we expect 
that the amplitude of the magnetic field should be reduced, 
compared to the case where $F_B$ is neglected  
and the magnetic field is not treated self-consistently 
(as in BdG calculations).

\section{Summary}
We have derived the GL equations and GL free energy for the 
flux phase and superconductivity microscopically from 
the two-dimensional $t-J$ model. 
The derived GL theory can be used to study
various problems in high-$T_c$ superconductivity, 
e.g., states near a surface or impurities, 
and the effect of an external magnetic field. 
The latter issue is important to distinguish 
theories proposed to explain  surface-state properties 
of cuprates.\cite{Lee2,Fogel2} 
Since the GL theory derived microscopically directly reflects 
the electronic structure of the system, e.g., the shape 
of the Fermi surface that changes with doping,   		
it can provide more useful information than that from  
phenomenological GL theories.\cite{cha2,Yu} 
In order to discuss the above mentioned problems, 
numerical calculations that treat magnetic fields 
as well as the OPs self-consistently 
are necessary, and we will examine them in a separate study. 

\begin{acknowledgment}
 The author thanks M. Hayashi and H. Yamase for useful discussions. 
\end{acknowledgment}

\appendix
\section{Coupling between Flux Phase Order Parameter and Magnetic Field}

In this appendix, we derive a term that couples 
the flux phase OP with the magnetic field, i.e., 
the last term in Eq.(19). 
This is the zeroth-order term in the GL equation for 
$\Pi$ (first-order term in the GL free energy $F_\Pi$), 
which arises from the substitution of 
${\tilde G}_0$ in Eq.(9) into  Eq.(13). 
The contribution to $Y_{j+{\hat x},j}$ is calculated as
\begin{eqnarray}
& \displaystyle 
iT \sum_{\varepsilon_n} \big[{\tilde G}_0(j+{\hat x},j,i\varepsilon_n)
- {\tilde G}_0(j,j+{\hat x},i\varepsilon_n) \big] \nonumber  \\
= & \displaystyle 
iT \sum_{\varepsilon_n} \frac{1}{N}\sum_p G_0({\bf p},i\varepsilon_n)
\big[e^{ip_x}e^{i\phi_{j+{\hat x},j}} 
- e^{-ip_x}e^{i\phi_{j,j+{\hat x}}} \big] \nonumber  \\
& \displaystyle 
\sim iT\sum_{\varepsilon_n} \frac{1}{N}\sum_p G_0({\bf p},i\varepsilon_n)
\Big[e^{ip_x}\Big(1- \frac{i\pi}{\phi_0}A_x({\bf r}_j+\frac{{\hat x}}{2}\Big)\Big)
- e^{-ip_x}\Big(1+ \frac{i\pi}{\phi_0}A_x\Big({\bf r}_j+\frac{{\hat x}}{2}\Big)\Big)
\Big] \nonumber \\ 
& \displaystyle 
=\frac{2\pi}{\phi_0} T\sum_{\varepsilon_n}\frac{1}{N}\sum_p 
\cos p_x G_0({\bf p},i\varepsilon_n)  A_x\Big({\bf r}_j+\frac{{\hat x}}{2}\Big)
\nonumber \\
& \displaystyle 
= \frac{2\pi}{\phi_0} \frac{1}{N} \sum_p \cos p_x  f(\xi_p) 
A_x\Big({\bf r}_j+\frac{{\hat x}}{2}\Big)
\nonumber \\
& \displaystyle 
= \frac{\pi\chi_0}{\phi_0} A_x\Big({\bf r}_j+\frac{{\hat x}}{2}\Big)
\end{eqnarray}
Contributions to $Y_{j+{\hat x}+{\hat y},j+{\hat x}}$, 
 $Y_{j+{\hat y},j+{\hat x}+{\hat y}}$, and  $Y_{j,j+{\hat y}}$ can be 
 calculated similarly. Substituting them into Eq.(16), the lowest-order term 
 is given by 
 \begin{eqnarray}
&  \displaystyle 
 \frac{\pi\chi_0}{4\phi_0} e^{i{\bf Q}\cdot{\bf r}_j} 
 \Big[ A_x\Big({\bf r}_j+\frac{{\hat x}}{2}\Big) 
 -  A_x\Big({\bf r}_j+\frac{{\hat x}}{2}+{\hat y}\Big) 
+  A_y\Big({\bf r}_j+{\hat x}+\frac{\hat y}{2}\Big)
- A_y\Big({\bf r}_j+\frac{\hat y}{2}\Big) \Big] 
\nonumber \\
& \displaystyle 
\sim \frac{\pi \chi_0}{4\phi_0} e^{i{\bf Q}\cdot{\bf r}_j} 
\Big[\partial_x A_y({\bf {\tilde r}}_j ) 
- \partial_y A_x({\bf {\tilde r}}_j )\Big]. 
 \end{eqnarray}
Here, we have expanded ${\bf A}$ around the center of a plaquette,  
 ${\bf {\tilde r}}_j = {\bf r}_j + {\hat x}/2 + {\hat y}/2$. 
By multiplying an appropriate factor, this gives the last term in 
Eq. (19).
Equation (A.2) shows that the flux phase OP, $\Pi$, which is 
defined at ${\bf {\tilde r}}_j$, 
couples to a magnetic field at the same point. 
In deriving the other terms of GL equations, we approximate 
$Y_{j+{\hat x}+{\hat y},j+{\hat x}} \sim -Y_{j-{\hat y},j}$ and 
$Y_{j+{\hat y},j+{\hat x}+{\hat y}} \sim Y_{j-{\hat x},j}$,  
assuming the slow variation of $|Y_{j,\ell}|$. 
Namely, in these terms, we approximate
\begin{eqnarray}
\Pi(j) \sim \frac{1}{4} (Y_{j+{\hat x},j} + Y_{j-{\hat x},j} 
- Y_{j+{\hat y},j} - Y_{j-{\hat y},j})e^{i{\bf Q}\cdot{\bf r}_j}.
\end{eqnarray}
This means that while the flux phase OP is defined at the center  
of a plaquette, ${\bf {\tilde r}}_j$, 
it couples to SCOPs defined at the neighboring site ${\bf r}_j$. 
If we use Eq. (A.3) to calculate the coupling between the 
flux phase OP and the vector potential, we would get 
a term of the form 
$\Pi(\partial_x A_x - \partial_y A_y)\cos ({\bf Q}\cdot{\bf r}) $ 
in $F_\Pi$, 
which is not gauge invariant and thus inappropriate.

\appendix
\section{Coefficients in GL Equations and GL Free Energy}
The coefficients appearing in the GL equations  
and GL free energy are given as follows:  
\begin{eqnarray}
& \displaystyle\alpha_{d(s)} = 3J \Big(1-\frac{3J}{4N}\sum_p 
I_1(p) \omega_{d(s)}^2(p) \Big), \\
& \displaystyle  \beta_{d(s)} = \frac{81J^4}{32N}\sum_p 
I_2(p) \omega_{d(s)}^4(p), \\
& \displaystyle  \gamma_1= \frac{81J^4}{8N}\sum_p 
I_2(p) \omega_d^2(p) \omega_s^2(p), 
\ \ \  \gamma_2 = \frac{1}{4} \gamma_1, \\
& \displaystyle  K_{d(s)}=  \frac{9J^2}{8N} \sum_p I_2(p) 
\Big(\frac{\partial \xi_p}{\partial p_x}\Big)^2 \omega_{d(s)}^2(p),  \\
& \displaystyle  K_{ds}= \frac{9J^2}{8N} \sum_p I_2(p) 
\Big(\frac{\partial \xi_p}{\partial p_x}\Big)^2 \omega_d(p) \omega_s(p), \\
& \displaystyle \alpha_\Pi = \frac{3J}{4}\Big[1 +\frac{3J}{4N}
\sum_p I_3(p)\omega_d^2(p)\Big], 
\\
& \displaystyle \beta_\Pi = \frac{1}{2}\Big(\frac{3J}{4}\Big)^4
\frac{1}{N}\sum_p I_4(p)\omega_d^4(p), 
\\
& \displaystyle K_\Pi =  -\frac{1}{2}\Big(\frac{3J}{4}\Big)^2
\frac{1}{N}\sum_p \frac{\partial \xi_p}{\partial p_x}
\frac{\partial \xi_{p+Q}}{\partial p_x}
I_4(p)\omega_d^2(p), 
\\
& \displaystyle \alpha_0 = -\frac{3\pi J}{8\phi_0}\chi_0, 
\\
& \displaystyle \gamma_{d\Pi} =  -4\Big(\frac{3J}{4}\Big) ^4
\frac{1}{N}\sum_p\Big[I_5(p)+2I_6(p)\Big]\omega_d^4(p), 
\\ 
& \displaystyle \gamma_{s\Pi} = -4\Big(\frac{3J}{4}\Big) ^4
\frac{1}{N}\sum_p\Big[I_5(p)+2I_6(p)\Big]\omega_d^2(p)\omega_s^2(p), 
\end{eqnarray}
where $\omega_d(p) = \cos p_x -\cos p_y$ and 
$\omega_s(p) =\cos p_x + \cos p_y$,
and the summation on $p$ is taken over the first Brillouin zone. 
The functions appearing in the integrals are defined as 
\begin{eqnarray}
I_1(p) = & \displaystyle 
T\sum_{\varepsilon_n} G_0(p,i\varepsilon_n)G_0(p,-i\varepsilon_n), \\
I_2(p) = & \displaystyle 
T\sum_{\varepsilon_n} G_0^2(p,i\varepsilon_n)G_0^2(p,-i\varepsilon_n), \\
I_3(p) = & \displaystyle 
T\sum_{\epsilon_n}G_0(p,i\varepsilon_n)G_0(p+Q,i\varepsilon_n), \\
I_4(p) = & \displaystyle 
T\sum_{\epsilon_n}G_0^2(p,i\varepsilon_n)G_0^2(p+Q,i\varepsilon_n), \\
I_5(p) = & \displaystyle 
T\sum_{\epsilon_n}G_0(p,i\varepsilon_n)G_0(p,-i\varepsilon_n)
G_0(p+Q,i\varepsilon_n)G_0(p+Q,-i\varepsilon_n), \\
I_6(p) = & \displaystyle T\sum_{\epsilon_n}
G_0^2(p,i\varepsilon_n)G_0(p,-i\varepsilon_n)G_0(p+Q,i\varepsilon_n).
\end{eqnarray}


\end{document}